\documentclass{ws-procs9x6}

\newcommand{\gsim}{\buildrel > \over {_\sim}}
\newcommand{\lsim}{\buildrel < \over {_\sim}}
\newcommand{\ie}{{\it i.e.}}
\newcommand{\eg}{{\it e.g.}}
\newcommand{\cf}{{\it cf.}}
\newcommand{\etal}{{\it et al.}}
\newcommand{\gev}{{\rm GeV}}
\newcommand{\as}{\alpha_s}

\newcommand{\lqcd}{\Lambda_{QCD}}

\newcommand{\eq}[1]{(\ref{#1})}

\newcommand{\beq}{\begin{equation}}
\newcommand{\eeq}{\end{equation}}
\newcommand{\beqa}{\begin{eqnarray}}
\newcommand{\eeqa}{\end{eqnarray}}

\newcommand{\PL}[3]{{\it Phys.~Lett.}~{\bf {#1}},~{#2}~({#3})}
\newcommand{\NP}[3]{{\it Nucl.~Phys.}~{\bf {#1}},~{#2}~({#3})}
\newcommand{\PR}[3]{{\it Phys.~Rev.}~{\bf {#1}},~{#2}~({#3})}
\newcommand{\PRL}[3]{{\it Phys.~Rev.~Lett.}~{\bf {#1}},~{#2}~({#3})}
\newcommand{\ZP}[3]{{\it Z.~Phys.}~{\bf {#1}},~{#2}~({#3})}

\begin{document}

\title{%
{\hbox to\hsize{\normalsize\hfil\rm NORDITA-2002-58 HE}}
\vskip 40pt
Duality in Semi-Exclusive Processes$^*$}

\author{Paul Hoyer$^\dagger$}

\address{Nordita \\
Blegdamsvej 17 \\ 
DK - 2100 Copenhagen, Denmark\\ 
E-mail: hoyer@nordita.dk}

\maketitle

\abstracts{
Bloom-Gilman duality relates parton distributions to nucleon form factors and thus constrains the dynamics of exclusive processes. The quark electric charge dependence implies that exclusive scattering is incoherent on the quarks even at high momentum transfers. Data on semi-exclusive meson production exceeds the duality prediction by more than an order of magnitude and violates quark helicity conservation. This suggests that the subprocess is dominated by soft `endpoint' contributions which obey dimensional scaling. The large transverse size of the subprocess may explain the absence of color transparency in fixed angle processes.}

\footnote[0]{$^*$Talk at the ``Workshop on Exclusive Processes at High Momentum Transfer'' at Jefferson Laboratory, Newport News, USA (May 2002). Transparencies including figures may be found at http://www.nordita.dk$/_{\tilde{}}$hoyer .} 

\footnote[0]{$^\dagger$On leave from the Department of Physical Sciences, University of Helsinki, Finland. Adjoint Senior Scientist, Helsinki Institute of Physics.}

\section{Bloom-Gilman Duality}

The remarkable relation between DIS $eN \to eX $ and exclusive resonance production $eN \to eN^*$ known as Bloom-Gilman duality\cite{BG} has been confirmed and extended by data from JLab\cite{CEBAF}. Empirically,
\beq\label{bgd}
\int\limits_{\delta x} d x F_2^{scaling}(x) \propto \frac{d\sigma}{dQ^2}(eN\to eN^*) \propto |F_{pN^*}(Q^2)|^2
\eeq
where $F_{pN^*}(Q^2)$ is the exclusive $p \to N^*$ electromagnetic form factor. The Bjorken variable $x=Q^2/(W^2+Q^2-M_N^2)$ is given by the photon virtuality $Q^2$ and the hadron mass $W= M_{N^*}$. On the lhs of \eq{bgd} the leading twist structure function $F_2^{scaling}(x)$ is integrated over an interval $\delta x$ covering the $N^*$ mass region. This semi-local duality relation is approximately satisfied for each nucleon resonance region including the Born term: $N^* = P_{11}(938),\ P_{33}(1232),\ S_{11}(1535)$ and $F_{15}(1680)$. The magnitude and $x$-dependence of the scaling structure function is thus related to the magnitude and $Q^2$-dependence of the $N^*$ electromagnetic form factors. The duality relation is
approximately satisfied even at low $Q^2$.

Bloom-Gilman duality means that the DIS scaling function is coded into the $N^*$ form factors. This is surprising because hard inclusive and exclusive processes are usually thought to be determined by separate parts of the nucleon wave function. The $F_2$ structure function is (at lowest order in $\as$) an incoherent sum, weighted by $e_q^2$, of inclusive quark distributions built from non-compact multiparton Fock states. The exclusive form factor on the other hand is believed to be governed by the wave function of the valence Fock state $|qqq\rangle$ of transverse size $1/Q$ \cite{lb}. The virtual photon then couples {\em coherently} to the valence quarks, implying a dependence on the quark electric charge of the form $(\sum_q e_q)^2$. Such a different dependence on $e_q$ of the two sides in Eq. \eq{bgd} contradicts the observed fact that duality is satisfied in a semilocal sense for both proton and neutron targets\cite{CEBAF}.

Data thus indicates that the Fock states in the nucleon electromagnetic form factors have size $\gg 1/Q$ so that the contribution of each quark to $|F_{pN^*}(Q^2)|^2$ is $ \propto e_q^2$. With the standard scaling laws $F_{pN^*}(Q^2) \propto 1/Q^4$ and $F_2^{scaling}(x) \propto (1-x)^3$ both sides of Eq. \eq{bgd} have the same $Q^2$-dependence. Duality and incoherent exclusive scattering will then hold at arbitrarily high $Q^2$. We shall find further evidence below that exclusive processes at large momentum transfer involve non-compact Fock states.

\section{Semi-Exclusive Processes}

Further information on the relation between inclusive and exclusive scattering can be obtained from generalizations of Bloom-Gilman duality. It was already observed that the spin\cite{bgds} and nuclear target $A$ dependence\cite{bgda} of the resonances agrees with that of the DIS scaling region as required by duality.

\begin{figure}[th]
\epsfxsize=7cm   
\centerline{\epsfbox{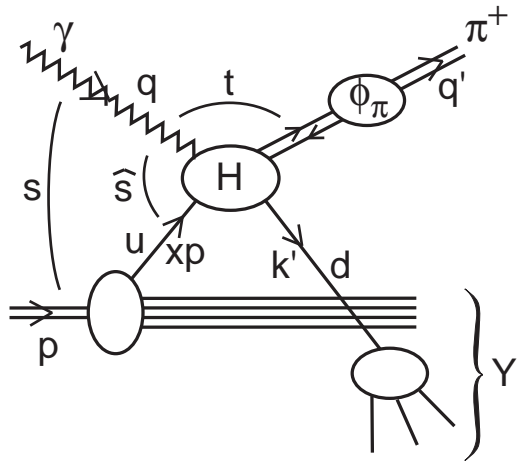}}
\caption{Semi-exclusive scattering. In the limit $\lqcd^2 \ll |t|, M_Y^2
\ll s$ the cross section factorizes into a hard subprocess cross section
$\hat\sigma(H)$ times a target parton distribution.}
\end{figure}

Semi-exclusive processes such as $\gamma p \to \pi^+ Y$ (Fig. 1) provide a qualitatively new testing ground of duality. In the kinematic limit where the total energy $s=(q+p)^2$ is much larger than
the mass of the inclusive system $Y$ ($s \gg M_Y^2 \gg \lqcd^2$) the produced $\pi^+$ meson is separated from the hadrons in $Y$ by a rapidity gap. When also the momentum transfer $t=(q-q')^2$ is large the $\pi^+$ is expected to be produced via a hard subprocess such as $\gamma u \to \pi^+ d$. We have then a generalisation of ordinary DIS, with the $eq \to eq$ subprocess replaced by $\gamma q \to \pi^+ q'$ and with the physical cross section given by\cite{BDHP} 
\beq \label{sec}
\frac{d\sigma}{dx\,dt}(\gamma p \to \pi^+ Y) = \sum_{q,q'} q(x)
\frac{d\sigma}{dt}(\gamma q \to \pi^+q')
\eeq
where $q(x)$ is the inclusive distribution of the struck quark in the target.
From the point of view of the target physics, there is a one-to-one
correspondence between the semi-exclusive process and ordinary DIS, with $Q^2
\leftrightarrow -t$ and $W^2 \leftrightarrow M_Y^2$. The momentum fraction of
the struck quark in the semi-exclusive process is thus $x = -t/(M_Y^2-t-M_N^2)$.

The close analogy with DIS makes it natural to study Bloom-Gilman duality for semi-exclusive processes\cite{ACW,ehk}. This links large momentum transfer exclusive cross sections to standard DIS structure functions and, via Eq. \eq{bgd}, to exclusive form factors, \eg,
\beq \label{sed}
\int_{\delta x}\left[u(x) + \bar{d}(x) \right] \frac{d\sigma}{dt}(\gamma u \to \pi^+d) \simeq \frac{d\sigma}{dt}(\gamma p \to \pi^+ N^*)
\eeq
With the standard behavior $u(x) \sim (1-x)^3$ we find, for $N^*=n$,
\beq
\frac{d\sigma}{dt}(\gamma p \to \pi^+ n) \propto \frac{1}{s^2 t^5}
\eeq
which is consistent with data\cite{rla}.

Combining Eqs. \eq{bgd} and \eq{sed} one obtains\cite{ehk}
\beq \label{dirrel}
\frac{d\sigma}{dt}(\gamma p \to \pi^+ n) =
16\pi^2\frac{\alpha\alpha_s^2 f_\pi^2}{|t|s^2}
\frac{G_{Mp}^2(-t)}{(1+r_{du}/4)}
\eeq
where $r_{du}$ is the $d/u$-quark distribution ratio for $x \to 1$. This estimate turns out to be nearly two orders of magnitude smaller than the measured\cite{rla} $\gamma p \to \pi^+ n$ cross section at $E_\gamma = 7.5\ \gev,\ |t| \simeq 2\ \gev^2$. The discrepancy is even worse at fixed angle -- in which case the $1/s^7$ scaling of both theory and data implies that the situation will not improve with momentum transfer. A similar result was found\cite{ehk} for Compton scattering $\gamma p \to \gamma p$, where duality underestimates data by about an order of magnitude.

Unfortunately there is no data on the semi-exclusive process $\gamma p \to \pi^+ Y$ in the continuum mass region of the inclusive system $Y$, where the prediction \eq{sec} could be tested directly. However, it is unlikely that the resonance/continuum ratio can be very different in the semi-exclusive process as compared to DIS. Hence the most likely reason for the failure of Eq. \eq{dirrel} is that \eq{sec} underestimates the semi-exclusive cross section. This could happen if endpoint contributions soften the meson production subprocess. In the previous section we saw that standard DIS duality also points in this direction.

\section{Spin Dependence of Semi-Exclusive $\rho$ Production} \label{rhoprod}

The ZEUS Collaboration recently published\cite{Zsemiex} striking data on polarization effects in semi-exclusive vector meson production $\gamma + p \to V + Y$, for $V=\rho^0, \phi, J/\psi$. The $\rho^0$ cross section scales as $d\sigma/dt \propto (-t)^n$, with $n = -3.21\pm 0.04 \pm 0.15$ in close agreement with $n= -3$ expected from dimensional counting for the subprocess $\gamma g \to q\bar q + g$. The corresponding data on $\phi$ production gives a power $n=-2.7\pm 0.1 \pm 0.2$. The ratio $\sigma(\phi)/\sigma(\rho) \simeq 2/9$ for $-t \gsim 3\ \gev^2$ is in accordance with flavor SU$_3$. These features suggest that the subprocess is hard and described by PQCD.

Helicity conservation at the level of the hard subprocess requires that the quark and antiquark are created with opposite helicities. Hence the $\rho$ meson they form is predicted to be longitudinally polarized, $\lambda_\rho =0$. At the level of the external particles, on the other hand, helicity conservation implies that the $\rho$ meson has the same helicity as the (real) projectile photon, \ie, $\lambda_\rho = \pm 1$. We thus have a situation where helicity conservation at the external particle level is in conflict with helicity conservation at the quark level.

The ZEUS data\cite{Zsemiex} shows that $s$-channel helicity is nearly conserved in the entire measured range ($-t \leq 6\ \gev^2$) for both $\rho$ and $\phi$ mesons.  Hence helicity is violated at the subprocess level. In PQCD this brings a suppression factor proportional to the quark mass squared, $m_q^2/(-t)$. The cross section is then expected to scale with a power $n=-4$, whereas the data is closer to the dimensional counting rule $n=3$. Taken at face value, this suggests that the subprocess is soft and endpoint dominated (one quark carrying most of the meson momentum) yet obeys the dimensional counting rule.
 
\section{Endpoint Behavior of Parton Subprocesses}

We have seen that Bloom-Gilman duality (in DIS as well as in semi-exclusive production) and quark helicity violation in vector meson production suggest a dominance of endpoint effects in exclusive hadron processes. PQCD estimates of the Lepage-Brodsky\cite{lb} hard exclusive scattering dynamics likewise show\cite{ilrk} that configurations where one quark carries most of the hadron momentum contribute importantly due to (nearly) on-shell internal propagators.

A possible reason for the failure to observe color transparency in large angle elastic $ep$\cite{ctep} and $pp$\cite{ctpp} scattering on nuclear targets is that the relevant nucleon Fock states are not compact, again due to endpoint contributions. We recently studied\cite{hltv} the size of the $\gamma u \to \pi^+ d$ subprocess at large momentum transfer $t$ in PQCD, based on the derivative of the cross section wrt. the virtuality $Q^2$ of the photon at $Q^2=0$. It turns out that while the amplitude itself is regular at the endpoints, $A\propto (e_u-e_d) \int dz\, \phi_\pi(z)/(1-z)$, the $Q^2$ derivative brings another factor of $1-z$ in the denominator. The integral over the quark momentum fraction $z$ is then singular at $z=1$, even though the pion distribution amplitude $\phi_\pi(z) \propto 1-z$. The singularity of the $Q^2$-derivative implies an {\em infinite} size for the photoproduction subprocess!

The situation for the quark helicity flip subprocess $\gamma g \to q\bar q + g$ (\cf\ section \ref{rhoprod}) is similar. The $\lambda_\rho =1$ amplitude has the form\cite{hltv}
\beq \label{flipamp}
A \propto \frac{m_q}{\sqrt{-t}} \int dz \frac{\phi_\rho(z)}{z^2(1-z)^2}\left[1+O\left(\frac{m_q^2}{t}\right)\right]
\eeq
and is thus endpoint sensitive. This may explain the dominance of the quark helicity flip amplitude in the ZEUS data\cite{Zsemiex}. Dimensional scaling can be understood by noting that the amplitude is not light-cone (LC) dominated at the endpoints: the soft quark moves with non-relativistic speed for $z \lsim \lqcd/\sqrt{-t}$. In this region the amplitude is not proportional to the LC distribution amplitude $\phi_\rho$ and there is no reason to expect that the numerator of the integrand in Eq. \eq{flipamp} vanishes. The linearly divergent integral then gives a factor $\sqrt{-t}/\lqcd$ which precisely compensates the $t$-dependence induced by the quark helicity flip. Details will be presented elsewhere\cite{hltv}.

\section{Conclusions}

There are several indications that large momentum transfer exclusive processes are not given by compact PQCD subprocesses and hadron distribution amplitudes. Rather, it appears that exclusive production is dominated by configurations where one quark carries most of the hadron momentum. The cross section scales dimensionally but does not obey quark helicity conservation nor color transparency.

\section*{Acknowledgments}
The results presented here were obtained in collaborations with Patrik Ed\'en, Alexander Khodjamirian, Jonathan Lenaghan, Kimmo Tuominen and Carsten Vogt. I am also grateful for discussions with Stan Brodsky and Jim Crittenden. Research supported in part by the European Commission under contract HPRN-CT-2000-00130.


\begin{thebibliography}{0}

\bibitem{BG}
E.~D.~Bloom and F.~J.~Gilman, \PRL{25}{1140}{1970} and
\PR{D4}{2901}{1971}.

\bibitem{CEBAF}
I.~Niculescu \etal, \PRL{85}{1182 and 1186}{2000}.

\bibitem{lb} G. P. Lepage and S. J. Brodsky, \PR{D22}{2157}{1980}.

\bibitem{bgds} A. Fantoni, Talk at the XL Int. Winter Meeting on Nuclear Physics, Bormio, Italy (January 2002), HERMES report 02-012 available at http://www-hermes.desy.de/notes/pub/doc-public-subject.html\#G1 .

\bibitem{bgda} J. Arrington, J. Crowder, R. Ent, C. Keppel and I. Niculescu, Jefferson Laboratory preprint PHY02-13 (2002).

\bibitem{BDHP}
S.~J.~Brodsky, M.~Diehl, P.~Hoyer and S.~Peigne,
\PL{B449}{306}{1999} [hep-ph/9812277].

\bibitem{ACW}
A.~Afanasev, C.~E.~Carlson and C.~Wahlquist,
\PR{D62}{074011}{2000} [hep-ph/0002271].

\bibitem{ehk} P. Ed\'en, P. Hoyer and A. Khodjamirian, {\it JHEP} {\bf 0110:040} (2001) [hep-ph/0110297].

\bibitem{rla} R.~L.~Anderson {\it et al.}, \PRL{30}{627}{1973};
\PR{D14}{679}{1976}.

\bibitem{Zsemiex} ZEUS Collaboration: S. Chekanov \etal, hep-ex/0205081.

\bibitem{ilrk} N. Isgur and C. H. Llewellyn Smith, \PRL{52}{1080}{1984} and \NP{B317}{526}{1989)};\\
A.V. Radyushkin \NP{A527}{153c}{1991} and \NP{A532}{141c}{1991};\\
J. Bolz, R. Jakob, P. Kroll, M. Bergmann and N.G. Stefanis, \ZP{C66}{267}{1995} [hep-ph/9405340].

\bibitem{ctep} T. G. O'Neill \etal, \PL{B351}{87}{1995} [hep-ph/9408260].

\bibitem{ctpp} A. Leksanov \etal, \PRL{87}{212301}{2001} [hep-ex/0104039].

\bibitem{hltv} P. Hoyer, J. T. Lenaghan, K. Tuominen and C. Vogt, in preparation.

\end{thebibliography}
\end{document}